\documentstyle[12pt,world_sci]{article}
\def\Wino{\boldmath $W\kern-.9em\hbox{\raise1.5ex\hbox{$\sim$}}_1$}
\def\Zino{\boldmath $Z\kern-.68em\hbox{\raise1.5ex\hbox{$\sim$}}_2$}
\def\wino{\hbox{$W\kern-.9em\hbox{\raise1.5ex\hbox{$\sim$}}_1$}}
\def\zinoa{\hbox{$Z\kern-.68em\hbox{\raise1.5ex\hbox{$\sim$}}_1$}}
\def\zinob{\hbox{$Z\kern-.68em\hbox{\raise1.5ex\hbox{$\sim$}}_2$}}

\def\mwino{$M_{W{\kern-.6em\hbox{\raise1.1ex\hbox{$\scriptstyle \sim$}}}_1}$}
\def\mgluino{$M_{\hbox{\scriptsize\it \~g}}$}
\def\msquark{$M_{\hbox{\scriptsize\it \~q}}$}
\def\mslepton{$M_{\hbox{\scriptsize\it \~l}}$}

\def\et{$E_t$}
\def\met{/\kern-.75em\hbox{$E_t$}}
\def\pbarp{$\overline{p}p$}

\def\nextline{\unskip\nobreak\hfill\break}
\def\abs#1{\mid\!{#1}\!\mid}

\begin{document}

\title{{\bf SEARCH FOR SUPERSYMMETRIC \Wino\ AND
\Zino\ STATES \\ USING THE D\O\ DETECTOR}}
\author{Susan K. Blessing\thanks{Representing the D\O\ Collaboration.}\\
{\em Department of Physics \\ Florida State University\\
Tallahassee, FL 32306, USA}}

\maketitle
\setlength{\baselineskip}{2.6ex}

\begin{center}
\parbox{13.0cm}
{\begin{center} ABSTRACT \end{center}
{\small \hspace*{0.3cm} The status of a search for the pair production
of the lightest chargino and second lightest neutralino states of the
minimal supersymmetric model is presented.  We have searched for four
tri-lepton final states: $eee$, $ee\mu$, $e\mu\mu$, and $\mu\mu\mu$, all
with missing transverse energy. }}
\end{center}

\section{Introduction}

%Supersymmetry (SUSY) is a theorized symmetry of nature
%relating bosons and fermions\cite{1}.
%Supersymmetric extensions of the Standard Model are attractive because they
%solve the so-called ``fine tuning problem" of the elementary scalar mass
%arising from radiative loop corrections and they unify the U(1), SU(2),
%and SU(3) couplings at $\sim\!\!10^{16}$\ GeV and are consistent with the
%LEP measurements\cite{2} of the running of the coupling constants.
%SUSY introduces an array of new
%particles; supergravity-inspired minimal supersymmetric standard models
%(MSSM) introduce the fewest new particles and
%provide guidance for mass relations and mixing among sparticles.

The search for signals from Supersymmetry\cite{1} at hadron colliders has
largely been centered on the search
for squark and gluino production, usually through events with large missing
transverse energy (\,\met).  A promising alternative at the Tevatron is the
search for production
of the lightest chargino (\wino) and next-to-lightest neutralino (\zinob),
with the assumptions that R-parity is conserved and the lightest neutralino
(\zinoa) is the lightest supersymmetric particle (LSP).
Although the mass limit established by LEP experiments\cite{2},
\mbox{\mwino $>$ 45 GeV/c$^2$}, excludes the on-shell process
\mbox{\pbarp\ $\rightarrow W^{\pm} \rightarrow$ \wino +\zinob, }the
off-shell process \mbox{$W^{\pm} \rightarrow$ \wino\zinob}\ can have a
sizeable rate due to large $W$\wino\zinob\ coupling\cite{3}.
The cross section for \wino\zinob\ production\cite{4} ranges from
$\sim\!\!10^3$\ pb to $\sim\!\!1$\ pb for \mwino\ between 40 and 100
GeV/c$^2$.
The final states of the \wino\ and \zinob\ are similar to the final states
of the $W^\pm$\ and $Z^0$ with the addition of~ \met\ from the LSP's.
The six possible event signatures are:
\vskip.1truein
\begin{table} [bth]\centering
\begin{tabular}{ l l}
2 jets + \met           &4 jets + \met \\
1 lepton + \met         &1 lepton + 2 jets + \met \\
2 leptons + 2 jets + \met {\phantom{aa}} \hfill &3 leptons + \met \\
\end{tabular}
\end{table}

\noindent The tri-lepton channel is the cleanest and most free of Standard
Model
backgrounds.  The branching fraction to three leptons, however, strongly
depends on the masses of the sleptons, sneutrinos, and squarks. In certain
supergravity inspired models\cite{5}, the
branching fraction to tri-lepton states may be enhanced, making this a
particularly attractive search channel.  In these models, mass
relations between the sparticle states lead to
\mbox{\mwino $\sim {1\over4}$\mgluino~-- ${1\over3}$\mgluino}. Thus
a search for \wino\ of mass up to 100 GeV/c$^2$ is comparable in reach
(in some subset of models) to searches for gluinos of 300 to 400 GeV/c$^2$.

We have searched for four final states: $eee$ + \met, $ee\mu$ + \met,
$e\mu\mu$ + \met, and \mbox{$\mu\mu\mu$ + \met}.
Studies using ISAJET\cite{6} show that the leptons are generally
centrally produced
and that the lepton with the lowest $p_t$\ is fairly soft.  Figure 1 shows
the distribution of $p_t$, $\eta$, and \met\ for \wino\zinob\ events.

\vskip2.5truein

\includegraphics{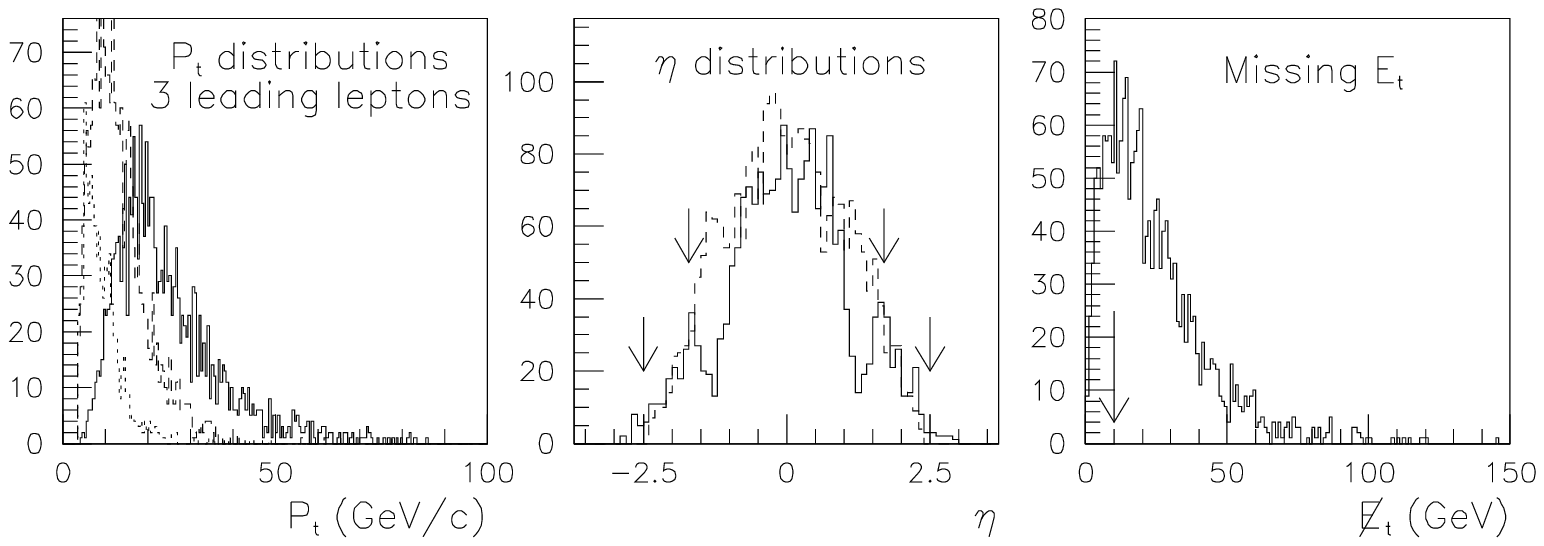}

{\small\noindent
Figure 1.  Transverse momentum distributions of the three leading leptons;
the $\eta$\ distribution of electrons (solid) and muons (dashed); and
the \met\
distribution for \wino\zinob\ ISAJET events after detector simulation
with \mwino = 65 GeV/c$^2$.  The arrows show analysis cut locations. }

\vskip.1truein
\section{The D\O\ Detector and Data Collection}

The data, based on a total integrated luminosity of
$13.8 \pm 1.6$\ pb$^{-1}$, were
collected during the 1992-1993 Tevatron run with the D\O\ Detector.
D\O\ is a general-purpose detector consisting of a non-magnetic central
tracking system, a finely segmented, nearly hermetic calorimeter, and
a toroidal muon spectrometer.  A detailed description of the detector can
be found elsewhere\cite{7}.

Six triggers were used in various combinations for the four channels:
\begin{itemize}
\item
1 muon with $p_t > 15$\ GeV/c, $\abs{\eta} < 1.7$
\item
1 muon with $p_t > 15$\ GeV/c, $\abs{\eta} < 1.7$ and \nextline
2 muons with $p_t > 10$\ GeV/c, $\abs{\eta} < 1.7$
\item
2 muons with $p_t > 3$\ GeV/c, $\abs{\eta} < 1.7$
\item
1 isolated EM object with $E_t > 20$\ GeV
\item
2 isolated EM objects with $E_t > 15$\ GeV
\item
1 EM object with  $E_t > 7$\ GeV and \nextline
1 muon with $p_t > 5$\ GeV/c, $\abs{\eta} < 1.7$
\end{itemize}

\vskip.1truein
\section{Data Analysis}

All channels were subject to offline trigger verification and required
to have $N_e + N_\mu = 3$\ and~ \met $\ge\! 10$\ GeV.
Muons were required to have $p_t(\mu) > 5$\ GeV/c, $\abs{\eta_\mu} < 1.7$,
and a matching minimum ionizing energy deposition in the calorimeter.
Electrons were required to have $E_t(e) > 5$\ GeV and $\abs{\eta_e} < 2.5$
and were subject to
calorimeter shape cuts.  Both electrons and muons were required to be
isolated.  To retain efficiency, electron and muon
identification was not as tight as possible, particularly for low energy
electrons and the third muon in the $\mu\mu\mu$\ channel.  After applying
the selection criteria, one event remains, an $ee\mu$\ event.
Both electrons in the event have associated tracks in the central detector;
\et$_1 = 37.3\pm .7$\ GeV, \et$_2 = 7.9 \pm .3$\ GeV,
$\eta_1 = 2.0$\ and $\eta_2 = 1.9$.
The muon is confirmed by the
presence of a minimum ionizing track in the calorimeter;
$p_t = 14.5 \pm 2.9$\ GeV/c and $\eta = 1.1$.
The~ \met$ = 38.7 \pm 2.7$\ GeV.
The kinematics of this event
are not really consistent with that expected from
\wino\zinob\ decay since all of the leptons are forward.  The topology
is suggestive of the process $Z^0 \rightarrow \tau\tau + \gamma$,
$\tau\tau \rightarrow e\mu$, where the photon converts
%$\gamma \rightarrow e^+e^-$\
and fakes an electron.
The $dE/dx$\ of both of the electrons is consistent with the
$dE/dx$\ of a conversion pair.

\vskip.1truein

\subsection{Efficiency}

A preliminary detection efficiency for each of the four channels was
determined
using ISAJET generated events.  One thousand events for each of the four
channels and for five \wino\ masses were generated and processed through
detailed detector and trigger simulations and the standard D\O\
reconstruction program.
Further work on the systematic errors associated with low energy lepton
efficiency is necessary.  Preliminary
values of the efficiency range from $\sim\!\!3$\% in the $e\mu\mu$\
channel for
\mwino = 45 GeV/c$^2$\ to $\sim\!\!15$\% in the $eee$\ channel for
\mwino = 100 GeV/c$^2$.  The MSSM parameters used as input to ISAJET are
shown in the table:
\vskip.17truein
\begin{table} [bth]\centering
\begin{tabular}{|c|c|l|}\hline
Parameter    &Value &\multicolumn{1}{c|} {Decription}\\ \hline
tan$\beta$   &2.0      &Ratio of the Higgs vacuum \\
             &         &expectation values\\ \hline
$M_{H^+}$    &500 GeV  &Mass of the charged Higgs \\ \hline
\mslepton    &200 GeV  &Slepton mass \\ \hline
\msquark     &1000 GeV &Squark mass \\ \hline
$M_t$        &150 GeV  &Top quark mass \\ \hline
\mgluino     &160-355 GeV &Gluino mass \\ \hline
\end{tabular}
\end{table}
\vskip.1truein
\subsection{Background}

The largest source of background is due to misidentification of one or more
of the leptons, for example in Drell-Yan + jet production where the
jet is largely electromagnetic and is misidentified as an electron.
Sources of background from misidentification include QCD three jet events,
\mbox{Drell-Yan + $b$\ jets}, \mbox{Drell-Yan + jets},
\mbox{Drell-Yan + $\gamma$}, \mbox{$Z^0$ + jets}
where $Z^0 \rightarrow l \overline{l}$, and $W^\pm$ + 2 jets.  Physics
backgrounds include $W^\pm Z^0 \rightarrow$ 3 leptons,
\mbox{$b\overline{b}$ + $\ge$ 1 additional $b$ jet},
and \mbox{$b\overline{b}$ + $\ge 1\ c$\ jet}.  A conservative
estimate (using the highest preliminary efficiency) of the number of
background events is shown in the table:
\vskip.17truein
\begin{table} [bth]\centering
\begin{tabular}{|c|c|c|c|c|}\hline
                   &eee   &ee$\mu$  &e$\mu\mu$  &$\mu\mu\mu$ \\ \hline
Misidentification  &0.43  &0.13     &0.21       &0.05 \\ \hline
Physics            &0.35  &0.08     &0.07       &0.18  \\ \hline
Total              &0.78  &0.22     &0.28       &0.23  \\ \hline
\end{tabular}
\end{table}

%\subsection{Results}
%
%The measurement of a limit on $\sigma \times Br$\ depends on a good
%knowledge of the efficiency for detecting the signal and the systematic
%errors involved in the measurement.  Since this work is not yet complete,
%we do not present a limit.

\section{Summary}

We are making good progress in a search for
\wino\zinob $\rightarrow$ tri-leptons using the D\O\ Detector at the
Fermilab Tevatron.  The current study is based on an integrated luminosity
of $13.8 \pm 1.6$\ pb$^{-1}$ from the 1992-1993 Tevatron run.  During the
Tevatron run now in progress, we are using triggers optimized for the low
$p_t$
leptons in these events and, with an expected integrated luminosity
of 50 -- 100 pb$^{-1}$, anticipate inproving the upper limit on
$\sigma \times Br$\ by a factor of $\approx 3-5$.
\vskip.1truein
{ }

\bibliographystyle{unsrt}

\begin{thebibliography}{99.}
\bibitem{1} See, {\it e.g.}, H.P. Nilles, {\sl Phys. Rep.} {\bf 110}
(1984) 1;
P. Nath, R. Arnowitt, and A. Chamseddine, {\sl Applied N=1 Supergravity},
ICTP Series in Theoretical Physics, Vol. I, World Scientific (1984);
H. Haber and G. Kane, {\sl Phys. Rep.} {\bf 117} (1985) 75.
%\bibitem{2} U. Amaldi, W. de Boer and H. F\"{u}rstenau, {\sl Phys. Lett.}
%{\bf B260} (1991) 447; P. Langacker and M. Luo, {\sl Phys. Rev.} {\bf D44}
%(1991) 817; U. Amaldi, \emet, {\sl Phys. Lett.} {\bf B281} (1992) 374.
\bibitem{2} T. Medcalf (for the ALEPH Collaboration),
in {\sl International Workshop on Supersymmtry and
Unification of Fundamental Interactions, Boston, Northeastern Univ,
29 March 1993 -- 1 April 1993}, ed. Pran Nath  (World Scientific, 1993)
R. M. Brown (for the OPAL Collaboration), {\it ibid.};
B. Zhou (for the L3 Collaboration), {\it ibid.}
\bibitem{3} R. Arnowitt and P. Nath, {\sl Mod. Phys. Lett.} {\bf A2} (1987)
331; H. Baer and X. Tata, {\sl Phys. Rev.} {\bf D47} (1993) 2739;
H. Baer, C. Kao, and X. Tata, {\sl Phys. Rev.} {\bf D48} (1993) 5175.
\bibitem{4} H. Baer, private communication.
\bibitem{5} G. Ross and R. Roberts, {\sl Nucl. Phys.} {\bf B377} 571;
R. Arnowitt and P. Nath, {\sl Phys. Rev. Lett.} {\bf 69} (1992) 725;
S. Kelley \emet, {\sl Nucl. Phys.} {\bf B398} (1993) 3.
\bibitem{6} ISAJET V7.06, F. Paige and S. D. Protopopescu, BNL Report
No. 38304 (1986); H. Baer, F. Paige, S. D. Protopopescu, and X. Tata,
{\sl Simulating Supersymmetry with ISAJET 7.0/ISASUSY 1.0},
FSU-HEP-930329 (1993).
\bibitem{7} S.~Abachi \emet,  {\sl Nucl. Instr. and Meth.}
{\bf A338} (1994) 185.
\end{thebibliography}

\end{document}